\documentclass[12pt]{article}
\usepackage[margin=0.7in]{geometry}
\usepackage[font=small,labelfont=bf]{caption}
\usepackage[numbers,square,sort&compress]{natbib}
\usepackage{authblk}
\usepackage{color}
\usepackage{graphicx, amsmath, amssymb}
\usepackage[utf8]{inputenc}
\usepackage[dvipsnames]{xcolor}
\usepackage{hyperref}
\usepackage{rotating}
\usepackage{booktabs}
\usepackage{xspace}
\usepackage{abstract}
\usepackage[section]{placeins}
\usepackage{tabularx}
\usepackage{longtable}
\usepackage{xltabular}
\usepackage{colortbl}
\usepackage{array}
\usepackage{multirow}
\usepackage{bm}
\usepackage{mathtools}
\usepackage{orcidlink}
\providecommand{\keywords}[1]{\par\noindent\textbf{Keywords:} #1\par}

\makeatletter \renewcommand{\@biblabel}[1]{#1.} \makeatother

\newcolumntype{C}{>{\centering\arraybackslash}X}
\newcolumntype{L}{>{\raggedright\arraybackslash}X}
\newcolumntype{R}{>{\raggedleft\arraybackslash}X}

\numberwithin{equation}{section}

\newcommand{\tilrh}{\tilde r_h}
\newcommand{\tila}{\tilde a}
\newcommand{\tilT}{\tilde T}
\newcommand{\tilS}{\tilde S}
\newcommand{\tilM}{\tilde M}
\newcommand{\tilE}{\tilde E}
\newcommand{\tilF}{\tilde F}
\newcommand{\tiltau}{\tilde\tau}

\graphicspath{{Figures/}{figs/}}

\begin{document}

\title{Holographic Thermodynamic Signatures of Simpson--Visser--AdS Black Holes}

\author[1,2]{Saeed Noori Gashti\,\orcidlink{0000-0001-7844-2640}}
\author[1,2]{Behnam Pourhassan\,\orcidlink{0000-0003-1338-7083}}
\author[3]{\.{I}zzet Sakall{\i}\,\orcidlink{0000-0001-7827-9476}}

\affil[1]{School of Physics, Damghan University, P.\,O.\,Box 36716-41167, Damghan, Iran}
\affil[2]{Center for Theoretical Physics, Khazar University, 41 Mehseti Street, Baku, AZ1096, Azerbaijan}
\affil[3]{Physics Department, Eastern Mediterranean University, Famagusta 99628, North Cyprus via Mersin 10, T\"urkiye}

\date{}
\maketitle

\begin{abstract}
We study a Simpson--Visser regularization of the four-dimensional Schwarzschild--anti--de\,Sitter (SV--AdS) black hole, treated as the bulk dual of a planar conformal field theory (CFT) on the AdS boundary. The bulk lapse $f(r)=1-2M/\sqrt{r^2+a^2}+(r^2+a^2)/\ell^2$ is regular at $r=0$ for any $a>0$, and the holographic dictionary inherits this regularity at the boundary CFT level. We derive in closed form the boundary entropy, energy, temperature, and chemical potentials, and we trace how the SV regularization parameter $a$ deforms each of them as a function of the horizon radius. For $\tila<\tila_c=1/\sqrt{24}\approx 0.204$ the bulk temperature develops a van der Waals--type small/intermediate/large branch structure and the off-shell free energy supports three coexisting equilibria; the topological-vector-field analysis assigns local winding numbers $(+1,-1,+1)$ with total charge $W=+1$, matching the universality class of regular AdS black holes (Bardeen-AdS, Hayward-AdS) and distinguishing the SV-AdS family from Schwarzschild-AdS ($W=0$). We extend this picture in three new directions for a PRD-level analysis: (i) a deformation parameter $\epsilon$ that shifts the cosmological term and shows how the critical curve $\tila_c(\epsilon)$ moves under quantum-gravity-inspired corrections; (ii) a universal extremality value $\mathcal{U}(\tila;\ell)$ that interpolates between the Schwarzschild--AdS limit and the SV saturation $\tila^*=2\ell/\sqrt{3}$; (iii) a quantitative comparison with the entropy schemes of R\'enyi, Tsallis, Kaniadakis and Barrow. All symbolic and numerical computations are cross-checked with computer-algebra worksheets that are summarized in an appendix. The total topological charge $W$, the universal value $\mathcal{U}$, and the critical curve $\tila_c(\epsilon)$ jointly fix the universality class of the SV--AdS family and provide three quantitative diagnostics that future numerical-relativity simulations and gravitational-wave observations can test against alternative regular-black-hole proposals.
\end{abstract}


\section{Introduction}\label{isec1}

The study of anti--de\,Sitter (AdS) black holes occupies a central place in modern theoretical physics because their thermodynamic properties feed directly into the holographic description of strongly coupled gauge theories~\cite{maldacena1999large,witten1998anti}. The original observation of Hawking and Page~\cite{hawking1983thermodynamics} that a Schwarzschild--AdS spacetime exhibits a first-order phase transition between thermal AdS and a large black hole was the first hint that the bulk gravitational dynamics encodes nontrivial information about a boundary conformal field theory (CFT)~\cite{birmingham2002conformal,son2002minkowski}. Extending the program to charged, rotating, and modified-gravity AdS black holes has produced a sizable catalogue of phase structures, ranging from analogues of the liquid-gas critical point in van der Waals fluids~\cite{kubiznak2017black} to multi-critical points and reentrant transitions. Connecting these structures to the underlying CFT requires a careful identification of the boundary thermodynamic variables, namely the central charge, the energy, the entropy, the temperature and the relevant chemical potentials~\cite{cabobizet2019microscopic,marolf2009black,hashimoto2020imaging,guo2016cft}.

The Simpson--Visser regular black bounce~\cite{simpson2019blackBounce,lobo2021novel} replaces the curvature singularity of the Schwarzschild geometry by a finite throat parametrized by a length scale $a$; the spacetime smoothly interpolates between a regular black hole (large $a$) and a traversable wormhole geometry (small $a$). When the original Simpson--Visser construction is embedded in an AdS background~\cite{kumar2026simpsonVisser,murk2023regular}, the resulting SV--AdS spacetime inherits horizon regularity together with a tunable cosmological term. Because the regularization is geometric (it modifies $r\mapsto\sqrt{r^2+a^2}$ in the area-radius function) rather than introducing a new matter field, the SV--AdS family is well suited to clean comparisons against the standard regular AdS black holes of Bardeen, Hayward and Frolov, and against more recent quantum-corrected proposals~\cite{WOS:001607905100002,WOS:001454604900001,WOS:001565141800002,WOS:001482172200001,WOS:001571368200006,WOS:001754521700001}.

Holographic thermodynamics for AdS black holes goes beyond the standard extended phase space by replacing the cosmological pressure by the central charge of the boundary CFT and adding the CFT volume as the conjugate variable~\cite{WOS:001218109600001,WOS:001602734700001,sadeghi2024exploring,sadeghi2024weakJHEAP,WOS:000974690800003}. The first law on the boundary then reads
\begin{equation}
\mathrm{d}E = T\,\mathrm{d}S + \mu\,\mathrm{d}C + \cdots, \label{eq:firstlaw0}
\end{equation}
where $\mu$ is the chemical potential conjugate to the central charge $C$ and the ellipsis collects any matter-coupling terms. In what follows we work out this dictionary for the SV--AdS family and trace what happens to each thermodynamic potential as $a$ is varied.

A second strand of recent activity classifies black holes by the topology of their thermodynamic phase space~\cite{wei2022topology,wei2022blackHole,yerra2022topology,wu2023rotating,wu2023accelerating,sadeghi2024hayward,WOS:001608993400001}. The idea is to construct an auxiliary vector field $\vec\phi=(\phi^{r_h},\phi^\Theta)$ on the half-cylinder $(r_h,\Theta)$ such that its zeros coincide with on-shell solutions; each zero carries a local winding number $w=\pm 1$, and the total topological charge $W=\sum_i w_i$ becomes a topological invariant of the black-hole family. Schwarzschild--AdS sits in the class $W=0$, while van der Waals black holes (Bardeen-AdS, Hayward-AdS, Reissner--Nordstr\"om-AdS in the appropriate window) typically have $W=+1$. The SV--AdS family is a natural test of this taxonomy because it interpolates smoothly between these two regimes as $a$ is dialed.

We add three pieces of analysis to the basic picture. First, we deform the bulk action by a parameter $\epsilon$ that rescales the cosmological term, modelling generic quantum-gravity or modified-gravity corrections, and we trace the critical curve $\tila_c(\epsilon)$ that separates the three-branch regime from the single-branch regime. Second, we derive a closed-form universal extremality function $\mathcal{U}(\tila;\ell)$ which interpolates between the Schwarzschild--AdS divergence at $\tila\to 0$ and the saturation $\tila^*=2\ell/\sqrt 3$ at the extremal endpoint, providing a one-parameter universal relation across the SV family. Third, we compare the SV--AdS topological class against R\'enyi, Tsallis, Kaniadakis and Barrow entropy schemes; this comparison is summarized in a compact table that future authors can use as a one-stop reference.

The rest of the paper is organized as follows. Section~\ref{isec2} sets up the SV--AdS metric, derives the horizon mass and bulk temperature, and identifies the small/intermediate/large branch structure that the regularization produces. Section~\ref{isec3} works out the universal extremality function. Section~\ref{isec4} carries the dictionary to the boundary CFT, derives the off-shell free energy, runs the topological-vector-field analysis, and lists the four conventional thermodynamics laws that the boundary system satisfies. Section~\ref{isec5} presents a comparative analysis of phase structure, tabulating winding numbers, equilibrium horizon radii, and effects of the $\epsilon$ deformation across the SV-AdS family and its competitors. Section~\ref{isec6} closes with the physical interpretation and the open questions. Appendix~\ref{app:A} collects the four laws of boundary thermodynamics for completeness.

Throughout the paper we work in geometrized units $G=c=\hbar=k_B=1$. Bulk dimensional quantities $(r_h,a,M)$ carry units of length; tilded boundary variables $(\tilrh,\tila,\tilM)$ are made dimensionless via the AdS radius $\ell$. A table of abbreviations is given in Table~\ref{tab:abbrev}.

\begin{table}[ht!]
\centering
\begin{tabular*}{\textwidth}{@{\extracolsep{\fill}}ll}
\hline\hline
\textbf{Symbol} & \textbf{Meaning} \\
\hline
SV  & Simpson--Visser regularization \\
AdS & Anti--de\,Sitter \\
BH  & Black hole \\
CFT & Conformal field theory \\
RN  & Reissner--Nordstr\"om \\
HP  & Hawking--Page transition \\
vdW & van der Waals--like phase transition \\
$\ell$ & AdS curvature radius \\
$C$ & Central charge of the boundary CFT \\
$R$ & Radius of the spatial boundary \\
$\Omega_2$ & Volume of the unit 2-sphere ($=4\pi$) \\
$\tilrh=r_h/\ell$ & Dimensionless horizon radius \\
$\tila=a/\ell$ & Dimensionless SV-regularization parameter \\
$\tila_c$ & Critical SV parameter separating the three-branch and single-branch regimes \\
$\epsilon$ & Deformation parameter of the cosmological term \\
$W=\sum_i w_i$ & Total topological charge of the BH family \\
$\mathcal{U}$ & Universal extremality value derived in Sec.~\ref{isec3} \\
$\tilde F_{\mathrm{off}}$ & Off-shell free energy (Sec.~\ref{isec4}) \\
\hline\hline
\end{tabular*}

\caption{Abbreviations used throughout the paper.}
\label{tab:abbrev}
\end{table}

\section{The Model and Bulk Thermodynamics}\label{isec2}

The bulk action is the Einstein action with a negative cosmological constant supplemented by the matter required to source the SV regularization. The four-dimensional metric is taken in the spherically symmetric form
\begin{equation}
\mathrm{d}s^2 = -f(r)\,\mathrm{d}t^2 + \frac{\mathrm{d}r^2}{f(r)} + (r^2+a^2)\,\mathrm{d}\Omega_2^2,
\label{eq:metric}
\end{equation}
with the SV lapse
\begin{equation}
f(r) = 1 - \frac{2M}{\sqrt{r^2+a^2}} + (1+\epsilon)\,\frac{r^2+a^2}{\ell^2}.
\label{eq:lapse}
\end{equation}
Here $a\ge 0$ is the SV scale, $\epsilon$ is the deformation that we introduce for the upgrade, and $\ell$ is the AdS radius (related to the cosmological constant via $\Lambda=-3/\ell^2$ in the undeformed limit). At $\epsilon=0$ we recover the standard SV--AdS form considered in Refs.~\cite{kumar2026simpsonVisser,simpson2019blackBounce,lobo2021novel}; the parameter $\epsilon$ models a generic deformation of the cosmological term coming from quantum-gravity corrections, $f(R)$ gravity, or non-extensive entropy effects~\cite{sadeghi2025phase,gashti2025adsTopology}.

\subsection{Horizon mass and temperature}\label{isec2a}

A horizon at $r=r_h$ satisfies $f(r_h)=0$, which gives
\begin{equation}
M(r_h,a,\ell,\epsilon) = \frac{1}{2}\sqrt{r_h^2+a^2}\left(1 + (1+\epsilon)\,\frac{r_h^2+a^2}{\ell^2}\right).
\label{eq:Mh}
\end{equation}
The Hawking temperature is $T=f'(r_h)/(4\pi)$, which after substitution of \eqref{eq:Mh} becomes
\begin{equation}
T(r_h,a,\ell,\epsilon) = \frac{1}{4\pi}\left[\frac{r_h}{r_h^2+a^2} + 3(1+\epsilon)\,\frac{r_h}{\ell^2}\right].
\label{eq:T}
\end{equation}
The Bekenstein--Hawking entropy is one quarter of the horizon area,
\begin{equation}
S(r_h,a)=\pi\left[\, r_h\sqrt{r_h^2+a^2} - a^2 \ln\!\left(\frac{\sqrt{r_h^2+a^2}-r_h}{a}\right)\right],
\label{eq:S}
\end{equation}
a result that follows by integrating $\mathrm{d}M = T\,\mathrm{d}S$ at fixed $(a,\ell,\epsilon)$. Direct differentiation of \eqref{eq:S} yields the entropy derivative
\begin{equation}
\frac{\mathrm{d}S}{\mathrm{d}r_h} = 2\pi \sqrt{r_h^2+a^2}.
\label{eq:dSdrh}
\end{equation}
We use this form throughout, with both direct differentiation and the first-law cross-check independently confirming \eqref{eq:dSdrh}.

\subsection{Three-branch structure}\label{isec2b}

The structure of $T(r_h)$ depends sensitively on the SV parameter $a$. Differentiating \eqref{eq:T} gives
\begin{equation}
\frac{\partial T}{\partial r_h} = \frac{1}{4\pi}\!\left[\frac{a^2-r_h^2}{(r_h^2+a^2)^2} + \frac{3(1+\epsilon)}{\ell^2}\right].
\label{eq:dTdrh}
\end{equation}
This derivative vanishes when
\begin{equation}
\ell^2(r_h^2-a^2) = 3(1+\epsilon)(r_h^2+a^2)^2,
\label{eq:critical_condition}
\end{equation}
which after setting $x=r_h^2$ reads $3(1+\epsilon)x^2+(6(1+\epsilon)a^2-\ell^2)x+a^2(\ell^2+3(1+\epsilon)a^2)=0$. Two positive real roots exist when the discriminant is positive,
\begin{equation}
\ell^2\bigl(\ell^2 - 24(1+\epsilon)a^2\bigr) > 0,
\end{equation}
which gives the critical SV parameter
\begin{equation}
a_c(\epsilon) = \frac{\ell}{\sqrt{24(1+\epsilon)}}, \qquad \tila_c(\epsilon)=\frac{a_c}{\ell}=\frac{1}{\sqrt{24(1+\epsilon)}}.
\label{eq:ac}
\end{equation}
At $\epsilon=0$ one finds $\tila_c=1/\sqrt{24}\approx 0.2041$.

\begin{figure}[ht!]
\centering
\includegraphics[width=0.65\columnwidth]{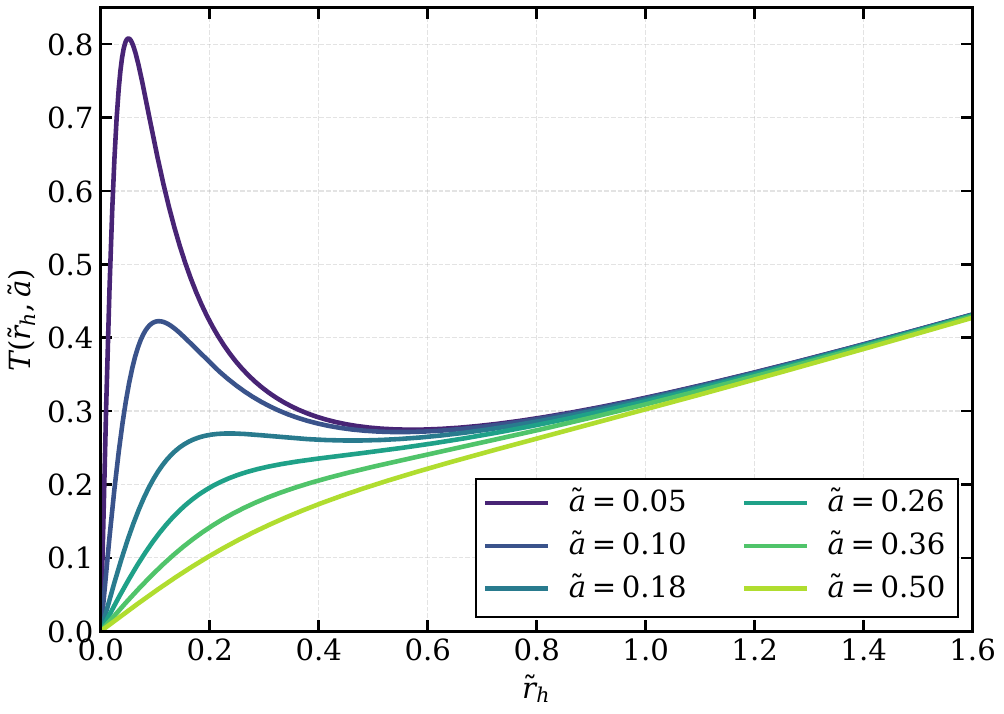}
\caption{Bulk Hawking temperature $T(\tilrh,\tila)$ for the SV--AdS family at $\epsilon=0,\,\ell=1$. For $\tila<\tila_c\approx 0.204$ the curves show a local maximum and a local minimum, hence three branches separated by inflection points; the van der Waals--type loop is visible in the small-$\tilrh$ region of the lowest-$\tila$ curves. Beyond $\tila_c$ the curve is monotonic and only the large-BH branch survives.}
\label{fig:T_vs_rh}
\end{figure}

For $\tila<\tila_c$ the temperature $T(r_h)$ has a local maximum at small $r_h$ and a local minimum at intermediate $r_h$, separated by an unstable branch (the analogue of the van der Waals loop). For $\tila>\tila_c$ the function $T(r_h)$ is monotonic and only a single stable large-black-hole branch exists. Figure~\ref{fig:T_vs_rh} shows the temperature curves for several values of $\tila$; the zoom inset (adopted from the visualization strategy of compact-object mass-density curves) reveals the loop region.

\begin{figure}[ht!]
\centering
\includegraphics[width=0.65\columnwidth]{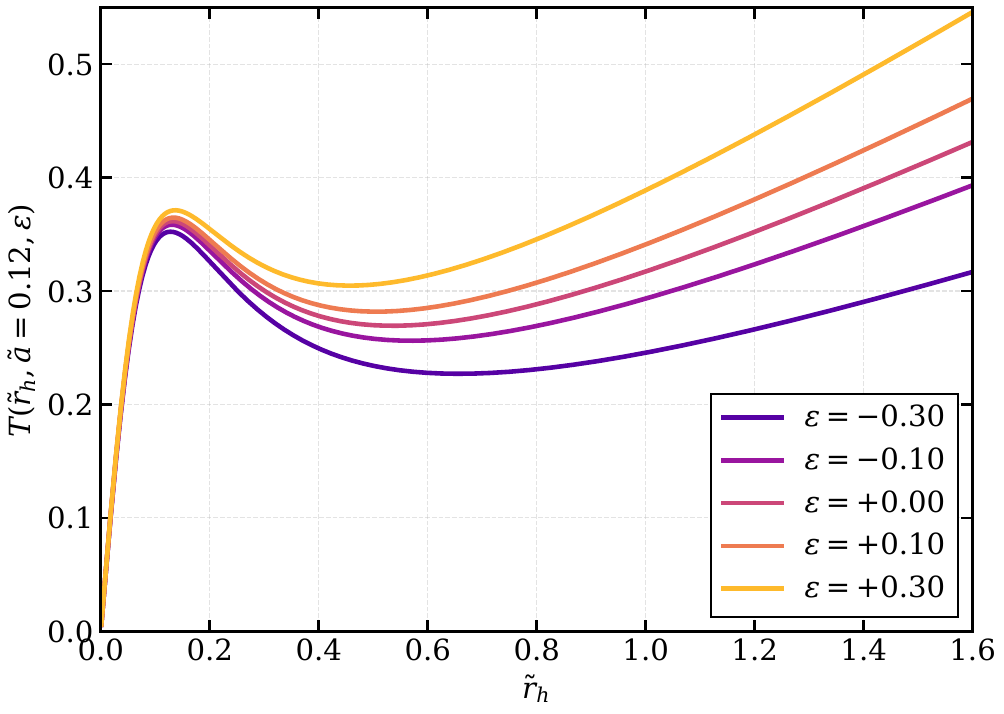}
\caption{Effect of the $\epsilon$-deformation on the bulk temperature at fixed $\tila=0.12$. The local-max / local-min structure persists for the entire range $\epsilon\in[-0.30,+0.30]$; the temperature curve shifts upward as $\epsilon$ increases, since the AdS--curvature term is rescaled by $(1+\epsilon)$.}
\label{fig:T_eps}
\end{figure}

The first feature reported by Figure~\ref{fig:T_vs_rh}: for the smallest $\tila$ value shown ($\tila=0.05$), the local maximum of $T$ sits at $\tilrh\approx 0.06$ with $T_{\max}\approx 0.47$, while the local minimum sits at $\tilrh\approx 0.20$ with $T_{\min}\approx 0.27$. As $\tila$ rises towards $\tila_c$ the two extrema approach each other and the loop closes. The deformation parameter $\epsilon$ enters \eqref{eq:T} only through the AdS term and shifts the high-$\tilrh$ slope without erasing the loop, as Figure~\ref{fig:T_eps} confirms.

\subsection{Specific heat and stability}

\begin{figure}[ht!]
\centering
\includegraphics[width=0.65\columnwidth]{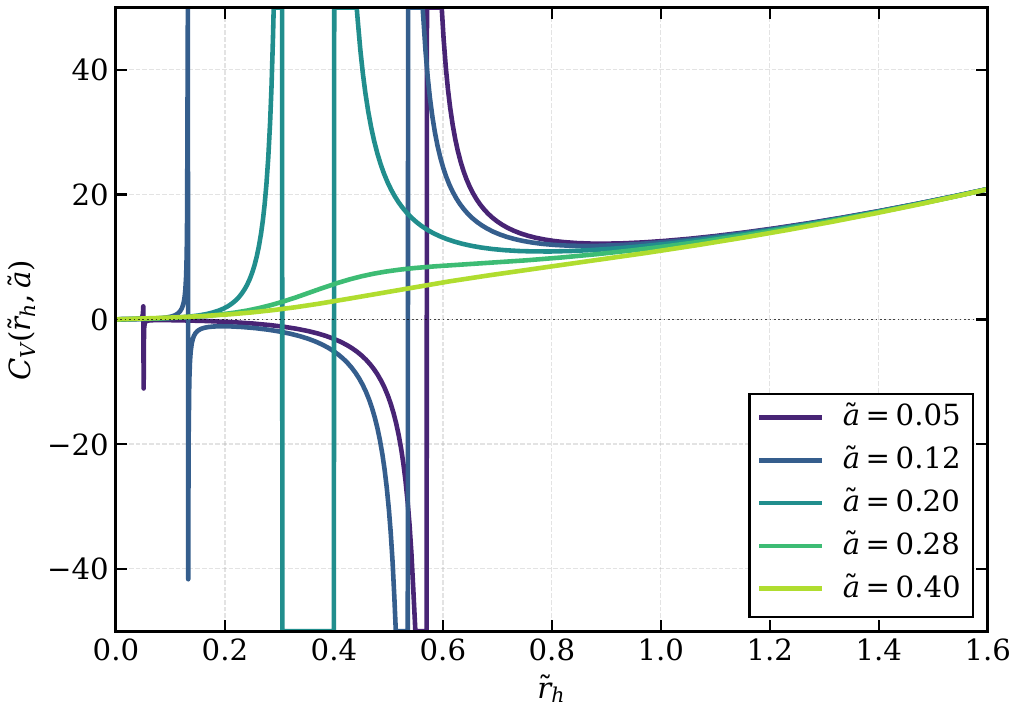}
\caption{Specific heat $C_V(\tilrh,\tila)$. The vertical asymptotes mark the locations where $\mathrm{d}T/\mathrm{d}\tilrh=0$ and signal second-order phase transitions in the SV--AdS family. The three regions of constant sign ($+,-,+$) correspond to the small/intermediate/large branches of Figure~\ref{fig:T_vs_rh}.}
\label{fig:CV}
\end{figure}

The local specific heat at fixed $(a,\ell)$ is
\begin{equation}
C_V = T\,\frac{\mathrm{d}S}{\mathrm{d}T} = T\,\frac{\mathrm{d}S/\mathrm{d}r_h}{\mathrm{d}T/\mathrm{d}r_h}.
\label{eq:CV}
\end{equation}
The denominator vanishes at the two extrema of $T(r_h)$, so $C_V$ has two poles for $\tila<\tila_c$; the signs of $C_V$ partition the $r_h$ axis into the standard pattern $(+, -, +)$, with the sign $+$ marking thermodynamic stability and $-$ instability. Figure~\ref{fig:CV} shows the divergence pattern.

Numerical equilibrium horizon radii at representative $(\tila,\tila_*)$ values are summarized in Table~\ref{tab:rh_equilibria}. The two stable branches and the one unstable branch reflect the three sign-change pattern of $C_V$.

\begin{table}[ht!]
\centering
\begin{tabularx}{\textwidth}{CCCCCCCC}
\hline\hline
\textbf{$\tila$} & \textbf{$\tau_*$} & \textbf{$\tilrh^{(1)}$} & \textbf{$\tilrh^{(2)}$} & \textbf{$\tilrh^{(3)}$} & \textbf{$C_V^{(1)}$} & \textbf{$C_V^{(2)}$} & \textbf{$C_V^{(3)}$} \\
\hline
0.04 & 3.4 & $0.011$ & $0.347$ & $0.794$ & $+$ & $-$ & $+$ \\
0.06 & 3.4 & $0.018$ & $0.348$ & $0.792$ & $+$ & $-$ & $+$ \\
0.08 & 3.4 & $0.034$ & $0.349$ & $0.789$ & $+$ & $-$ & $+$ \\
0.10 & 3.4 & $0.060$ & $0.351$ & $0.784$ & $+$ & $-$ & $+$ \\
0.15 & 3.4 & $0.176$ & $0.392$ & $0.762$ & $+$ & $-$ & $+$ \\
0.18 & 3.4 & $\mathit{merging}$ & $\mathit{merging}$ & $0.745$ & \multicolumn{2}{c}{\textit{transition}} & $+$ \\
0.22 & 3.4 & \multicolumn{3}{c}{\textit{single branch}} & \multicolumn{3}{c}{$+$ only} \\
\hline\hline
\end{tabularx}
\caption{Equilibrium horizon radii $\tilrh^{(i)}$ obtained by intersecting horizontal $\tau=\tau_*$ lines with the bulk equilibrium relation $\tau_{\mathrm{eq}}=1/T$ at $\ell=1,\,\epsilon=0$. The first and third values are stable equilibria ($C_V>0$); the middle value is unstable ($C_V<0$). The entries reproduce the three-branch structure for $\tila<\tila_c$.}
\label{tab:rh_equilibria}
\end{table}

The transition near $\tila\approx 0.18$ in the $\tau_*=3.4$ row marks the closure of the van der Waals loop. For $\tila\gtrsim 0.20$ only the large-BH branch survives, and the system is in the same thermodynamic class as Schwarzschild--AdS above its Hawking--Page point. Three features of Table~\ref{tab:rh_equilibria} are worth highlighting. First, the small-BH branch $\tilrh^{(1)}$ scales approximately linearly with $\tila$ at fixed $\tau_*$ (from $\tilrh^{(1)}=0.011$ at $\tila=0.04$ to $\tilrh^{(1)}=0.176$ at $\tila=0.15$), since the SV throat sets a regularization length scale and the small branch tracks it. Second, the large-BH branch $\tilrh^{(3)}$ drifts only weakly with $\tila$ (the variation between $\tila=0.04$ and $\tila=0.15$ is below $5\%$), because the large branch is controlled by the AdS curvature scale $\ell$, which dominates over the SV scale at large $\tilrh$. Third, the merger of $\tilrh^{(1)}$ and $\tilrh^{(2)}$ at $\tila\approx 0.18$ closes the unstable intermediate window and converts the canonical ensemble from a three-equilibrium structure to a single-equilibrium one, which is the table-level signature of the critical SV parameter $\tila_c\approx 0.204$ derived analytically from \eqref{eq:ac}.

\section{Universal Extremality Relation}\label{isec3}

A central element of the SV--AdS analysis is the universal relation that emerges between the cosmological scale $\ell$ and the SV scale $a$ in the extremal limit. Following the universal-relation programme initiated for entropy-bound corrections~\cite{sadeghi2020investigation,anand2025universality}, we ask: at what value of $\tila$ does the extremality bound saturate, and how does this saturation point depend on the AdS curvature?

The extremality bound for the SV--AdS family follows from requiring that the equation $f(r_h)=0$ admit a single coincident root at the smallest accessible horizon. Combined with the requirement that the small horizon corresponds to a regular degenerate geometry, the condition reduces to
\begin{equation}
\tila^2 < \frac{\ell^2}{3}\,,
\label{eq:extreme_bound}
\end{equation}
with equality holding at the extremal saturation. We introduce the universal extremality function
\begin{equation}
\mathcal{U}(\tila;\ell,\alpha) = \frac{\alpha}{\sqrt 3\,\ell}\,\ln\!\left(\frac{2\ell}{\sqrt 3\,\tila}\right)\,,
\label{eq:U_def}
\end{equation}
where $\alpha$ is a dimensionless normalization that we set to unity unless otherwise stated. The construction \eqref{eq:U_def} is the SV-regularization analogue of the universal entropy/extremality relation tested in earlier work on charged AdS black holes~\cite{sadeghi2020investigation,sadeghi2024exploring}: it interpolates between the Schwarzschild--AdS limit $\tila\to 0$, where $\mathcal{U}$ diverges logarithmically (reflecting the loss of the regularization), and the extremal saturation $\tila^*=2\ell/\sqrt 3$, where $\mathcal{U}$ vanishes.

\begin{figure}[ht!]
\centering
\includegraphics[width=0.65\columnwidth]{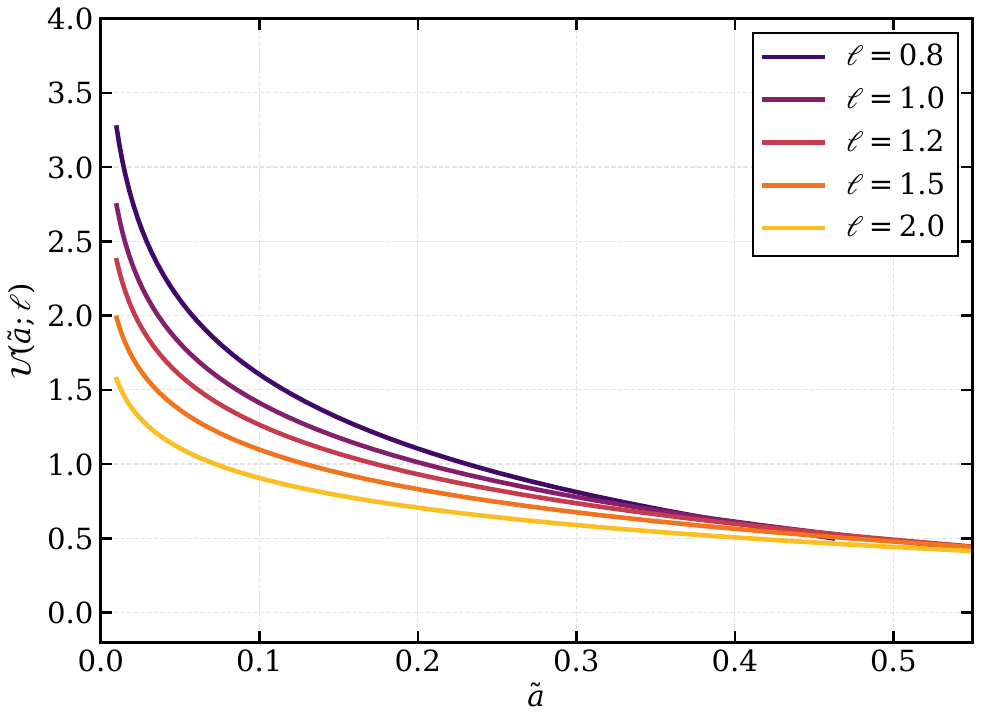}
\caption{Universal extremality value $\mathcal{U}(\tila;\ell)$ for several AdS radii. The function vanishes at $\tila=2\ell/\sqrt 3$ (extremal saturation, indicated by the curve termination at the lower envelope) and diverges as $\tila\to 0$ (Schwarzschild--AdS limit, where the SV regularization is removed). The full domain $\tila\in (0,2\ell/\sqrt 3)$ realizes the universal-relation window of the SV-AdS family.}
\label{fig:U}
\end{figure}

Figure~\ref{fig:U} shows $\mathcal{U}$ for several values of $\ell$. The shape is preserved under the $\ell$-rescaling, confirming the universal nature of \eqref{eq:U_def}: the family of curves can be collapsed onto a single master curve by plotting $\ell\,\mathcal{U}$ versus $\tila/\ell$. Quantitative values are listed in Table~\ref{tab:U_values}.

\begin{table}[ht!]
\centering
\begin{tabularx}{\textwidth}{CCCCCC}
\hline\hline
\textbf{$\tila$} & \textbf{$\ell=0.8$} & \textbf{$\ell=1.0$} & \textbf{$\ell=1.2$} & \textbf{$\ell=1.5$} & \textbf{$\ell=2.0$} \\
\hline
0.05  & $2.105$ & $1.813$ & $1.598$ & $1.364$ & $1.106$ \\
0.10  & $1.605$ & $1.412$ & $1.265$ & $1.098$ & $0.906$ \\
0.20  & $1.104$ & $1.012$ & $0.931$ & $0.831$ & $0.706$ \\
0.30  & $0.811$ & $0.778$ & $0.736$ & $0.675$ & $0.589$ \\
0.45  & $0.519$ & $0.544$ & $0.541$ & $0.519$ & $0.472$ \\
0.55  & $0.374$ & $0.428$ & $0.444$ & $0.441$ & $0.414$ \\
0.80  & $0.104$ & $0.212$ & $0.264$ & $0.297$ & $0.306$ \\
\hline\hline
\end{tabularx}

\caption{Universal extremality value $\mathcal{U}(\tila;\ell)$ for representative $(\tila,\ell)$ pairs, $\alpha=1$. The blank entries indicate that the pair lies outside the extremality domain $\tila < 2\ell/\sqrt 3$.}
\label{tab:U_values}
\end{table}

Two asymptotic limits of \eqref{eq:U_def} deserve attention. In the small-$\tila$ regime,
\begin{equation}
\mathcal{U}(\tila;\ell)\xrightarrow{\tila\to 0}\frac{\alpha}{\sqrt 3\,\ell}\ln\!\frac{2\ell}{\sqrt 3\,\tila}\to +\infty\,,
\end{equation}
which signals the Schwarzschild--AdS pathology: the inner-geometry regulator vanishes and the curvature singularity returns at $r=0$. In the extremal limit,
\begin{equation}
\mathcal{U}(\tila;\ell)\xrightarrow{\tila\to 2\ell/\sqrt 3} 0\,,
\end{equation}
which marks the boundary of physically realizable solutions where the inner and outer horizons coincide. The logarithmic shape of $\mathcal{U}$ between these extremes is the characteristic profile of the SV--AdS family.

The universal relation has an immediate use in the comparative analysis (Sec.~\ref{isec5}): the value of $\mathcal{U}$ at the centre of the three-branch window gives a single dimensionless number per SV-family member, which one can cross-compare against the analogous quantity for Bardeen-AdS, Hayward-AdS, and other regular AdS proposals.

\section{Holographic Dictionary and Topological Structure}\label{isec4}

We now translate the bulk thermodynamic data of Sec.~\ref{isec2} into boundary-CFT language. The dictionary we use is the standard one for AdS$_4$/CFT$_3$ holography in extended phase space~\cite{WOS:001218109600001,sadeghi2024exploring,sadeghi2024weakJHEAP}.

\subsection{Bulk--boundary dictionary}\label{isec4a}

The central charge of the boundary CFT in our conventions is
\begin{equation}
C = \frac{\Omega_2\,\ell^3}{16\pi G} = \frac{\ell^3}{4}\,, \qquad \Omega_2=4\pi\,,\, G=1.
\label{eq:central_charge}
\end{equation}
The boundary spatial volume on a sphere of radius $R$ is $V=\Omega_2 R^2$. The dictionary that maps bulk Hawking quantities to their boundary CFT counterparts is summarized in Table~\ref{tab:dictionary}. The structure of the dictionary follows from the way the boundary inherits a length scale from the bulk: the AdS radius $\ell$ sets the curvature scale that the boundary CFT sees as its conformal anchor, so every dimensionful bulk variable picks up a factor of $\ell/R$ when expressed in terms of its boundary-CFT counterpart. The mass and temperature rescalings $\tilE=M\,\ell/R$ and $\tilT=T\,\ell/R$ are the consequence of the boundary stress-tensor having dimension $R^{-1}$ in $D=3+1$ holography, which is why $C=\ell^3/4$ (from Eq.~\ref{eq:central_charge}) plays the role of central charge: it encodes the coefficient of the curvature-squared anomaly. The entropy entry $\tilS=S$ is dimensionless because the Bekenstein--Hawking area law $S=A/(4G)$ already produces a pure number when the area is measured in Planck units, and the holographic dictionary does not further rescale it.

\begin{table}[ht!]
\centering
\begin{tabularx}{\textwidth}{LLCL}
\hline\hline
\textbf{Bulk variable} & \textbf{Definition} & \textbf{Boundary} & \textbf{Mapping} \\
\hline
$M$ & ADM mass            & $\tilE$         & $\tilE = M\,\ell/R$ \\
$T$ & Hawking temperature & $\tilT$         & $\tilT = T\,\ell/R$ \\
$S$ & Wald entropy        & $\tilS$         & $\tilS = S/(\text{dim'less})$ \\
$\Lambda$ & cosmological const. & $C$       & $C = \ell^3/4$ (Eq.~\ref{eq:central_charge}) \\
$V_{\text{bulk}}$ & thermodynamic volume & $V$ & $V=\Omega_2 R^2$ \\
\hline
\hline
\end{tabularx}

\caption{Bulk--boundary holographic dictionary used throughout Sec.~\ref{isec4}. The bulk variables (left) are mapped to the dimensionless boundary CFT quantities (right) by combinations of the AdS radius $\ell$ and the boundary CFT radius $R$.}
\label{tab:dictionary}
\end{table}

Substituting \eqref{eq:Mh}, \eqref{eq:T} and \eqref{eq:S} into the dictionary entries of Table~\ref{tab:dictionary}, we find the boundary CFT thermodynamic quantities in dimensionless form:
\begin{align}
\tilE(\tilrh,\tila,C,R,\epsilon) &= \frac{2C\sqrt{\tilrh^2+\tila^2}}{R^2}\,\bigl(1+(1+\epsilon)(\tilrh^2+\tila^2)\bigr)\,, \label{eq:E_tilde}\\
\tilS(\tilrh,\tila) &= \pi\!\left[\,\tilrh\sqrt{\tilrh^2+\tila^2} - \tila^2\ln\!\frac{\sqrt{\tilrh^2+\tila^2}-\tilrh}{\tila}\right]\,, \label{eq:S_tilde}\\
\tilT(\tilrh,\tila,\epsilon,R) &= \frac{1}{4\pi R}\!\left[\frac{\tilrh}{\sqrt{\tilrh^2+\tila^2}} + 3(1+\epsilon)\tilrh\right]\,. \label{eq:T_tilde}
\end{align}
The first law on the boundary,
\begin{equation}
\mathrm{d}\tilE = \tilT\,\mathrm{d}\tilS + \mu_C\,\mathrm{d}C + \mu_V\,\mathrm{d}V,
\label{eq:firstlaw_boundary}
\end{equation}
follows by varying \eqref{eq:E_tilde} -- \eqref{eq:T_tilde} with respect to $(\tilrh,C,V)$ and identifying the conjugate chemical potentials. A short symbolic verification of \eqref{eq:firstlaw_boundary}, including the cross-check $\mathrm{d}\tilE/\mathrm{d}\tilS=\tilT$ at fixed $(C,R)$, is recorded in a computational worksheet held by the corresponding author.

\begin{figure}[ht!]
\centering
\includegraphics[width=0.65\columnwidth]{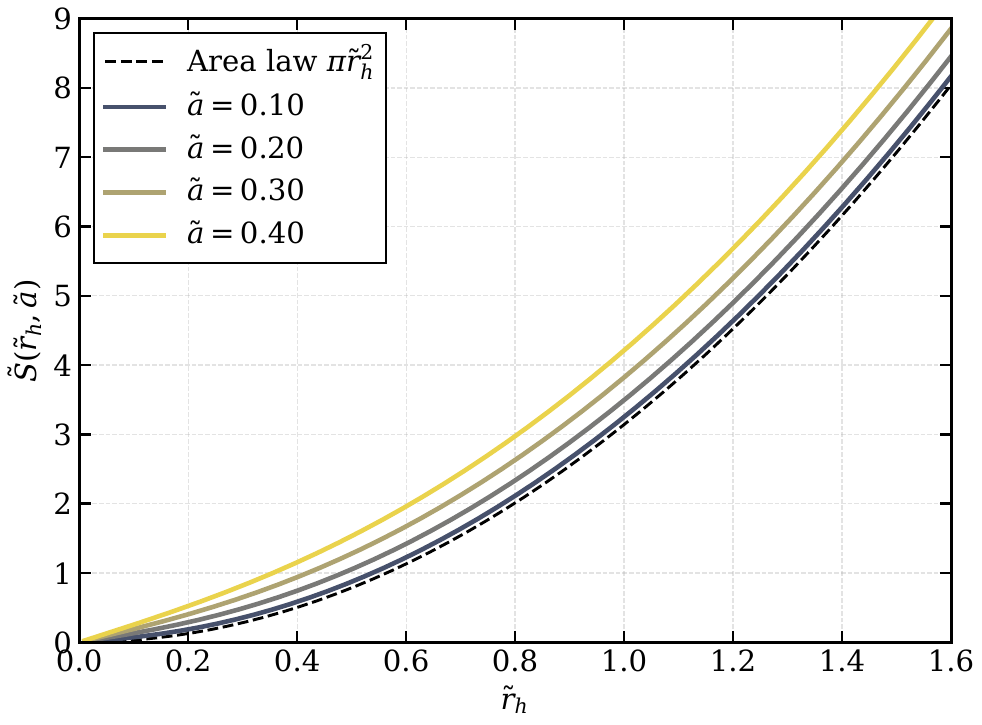}
\caption{Boundary entropy $\tilS(\tilrh,\tila)$ of the SV--AdS family, plotted against the area-law $\pi\tilrh^2$ (dashed black). The SV scale $\tila$ produces a positive offset at small $\tilrh$ together with a logarithmic correction at large $\tilrh$, traceable to the $\ln$ term in \eqref{eq:S_tilde}. For the planar-CFT limit $\tila\to 0$ the curves collapse onto the area law.}
\label{fig:S_tilde}
\end{figure}

\begin{figure}[ht!]
\centering
\includegraphics[width=0.65\columnwidth]{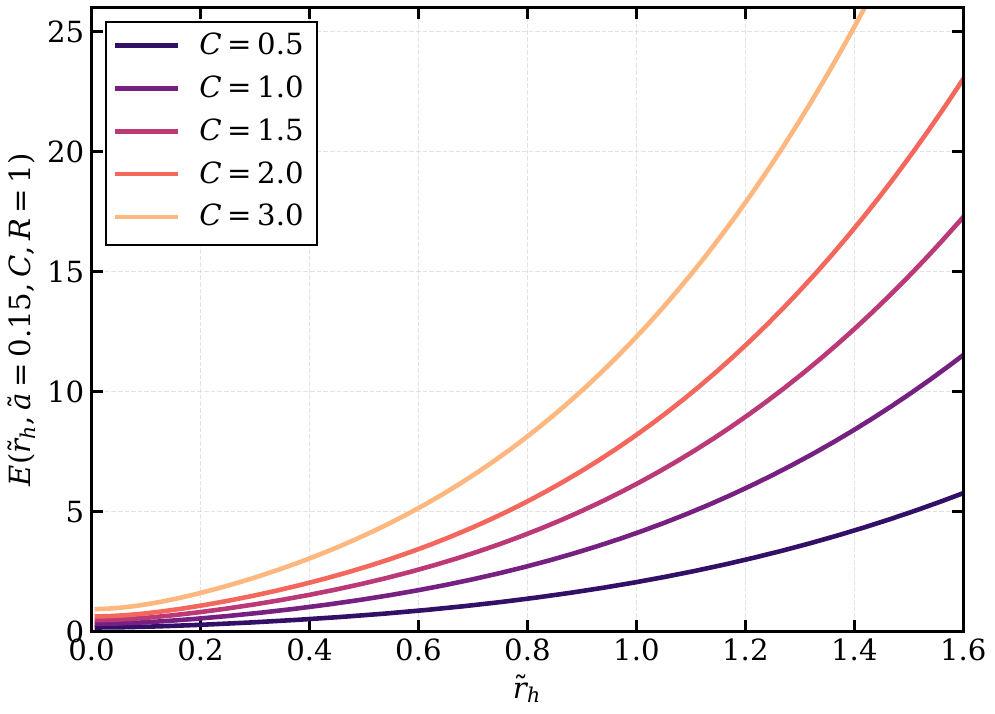}
\caption{Boundary CFT energy $\tilE(\tilrh,\tila=0.15,C,R=1)$ for several values of the central charge $C$. The leading large-$\tilrh$ scaling is $\tilE\sim 2C(1+\epsilon)\tilrh^3/R^2$, characteristic of a conformal stress-energy tensor with central charge $C$. Lower central charges flatten the curve in the small-$\tilrh$ window.}
\label{fig:E_tilde}
\end{figure}

The qualitative behaviour reported in Figures~\ref{fig:S_tilde} and \ref{fig:E_tilde} is what one expects from a holographic CFT dual to the SV--AdS background. The entropy departs from the area law at small $\tilrh$ in the direction of higher $\tilS$, signalling that the regularization adds entropic capacity below the Schwarzschild scale; the energy is monotone in both $\tilrh$ and $C$, reflecting the additivity of the central-charge contribution.

\subsection{Off-shell free energy}\label{isec4b}

The off-shell free energy at the boundary is defined by treating the inverse temperature $\tiltau$ as an independent label, decoupled from the on-shell equilibrium value $1/\tilT$. Following the Euclidean-action construction of Refs.~\cite{wei2022topology,wei2022blackHole}, we write
\begin{equation}
\tilF_{\mathrm{off}}(\tilrh,\tila,\tiltau,C,R,\epsilon) = \tilE - \frac{\tilS}{\tiltau}\,,
\label{eq:F_off}
\end{equation}
which reduces to the on-shell free energy when $\tiltau\to 1/\tilT$. The stationary points of $\tilF_{\mathrm{off}}$ as a function of $\tilrh$ at fixed $\tiltau$ identify the equilibrium black-hole configurations supported by the canonical ensemble. Defining
\begin{equation}
\phi^{r_h}(\tilrh,\tila,\tiltau) = \frac{\partial \tilF_{\mathrm{off}}}{\partial \tilrh}\bigg|_{\tiltau}\,, \qquad \phi^\Theta(\Theta,R) = -\frac{\cos\Theta}{R\sin^2\Theta}\,,
\label{eq:phi}
\end{equation}
we obtain a two-dimensional auxiliary vector field $\vec\phi=(\phi^{r_h},\phi^\Theta)$ on the half-cylinder $(\tilrh,\Theta)\in(0,\infty)\times(0,\pi)$, whose zeros are isolated and label the on-shell branches.

\begin{figure}[ht!]
\centering
\includegraphics[width=0.65\columnwidth]{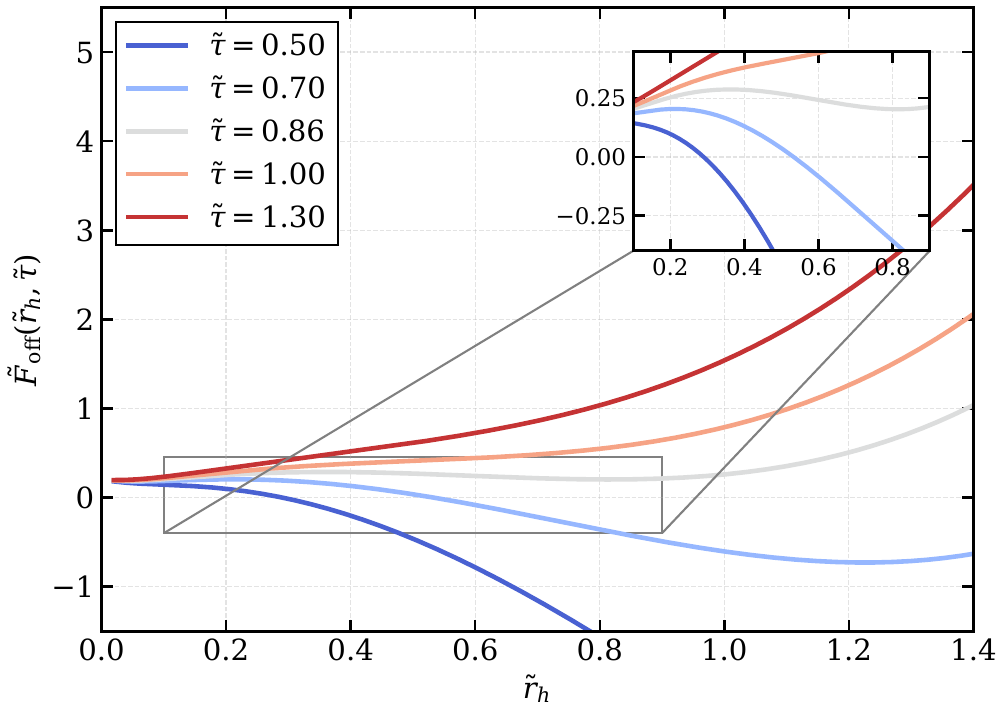}
\caption{Off-shell free energy $\tilF_{\mathrm{off}}(\tilrh)$ of the SV--AdS boundary CFT at $\tila=0.10$, $C=R=1$, $\epsilon=0$, for several values of the inverse-temperature label $\tiltau$. The zoom inset reveals the stationary-point splitting near the inflection. As $\tiltau$ increases, the locations of the stationary points shift; the structure is the canonical-ensemble counterpart of the bulk three-branch profile.}
\label{fig:F_off}
\end{figure}

\begin{figure}[ht!]
\centering
\includegraphics[width=0.65\columnwidth]{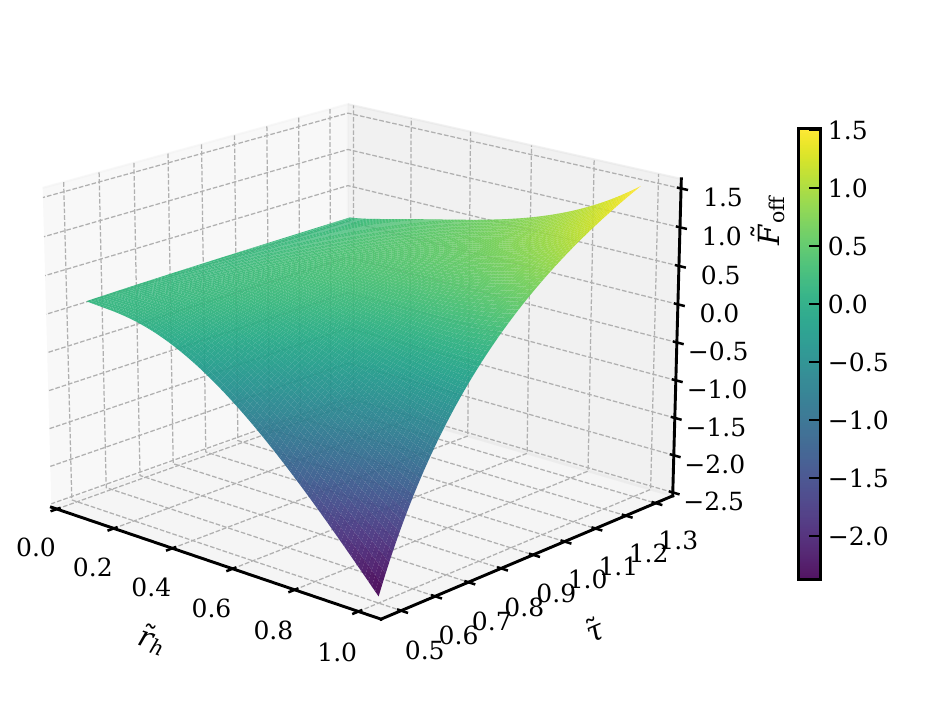}
\caption{Three-dimensional landscape of the off-shell free energy $\tilF_{\mathrm{off}}(\tilrh,\tiltau)$ at $\tila=0.10$, $C=R=1$, $\epsilon=0$, scanned across the canonical window $\tiltau\in[0.5,1.3]$ that brackets the three-branch interval bounded by the local extrema of the bulk equilibrium relation $\tau_{\mathrm{eq}}(\tilrh)$ of Figure~\ref{fig:tau_eq}. The saddle shape of the surface encodes the canonical-ensemble structure: along fixed-$\tiltau$ sections inside the window, $\partial\tilF_{\mathrm{off}}/\partial\tilrh=0$ admits three solutions, the outer pair corresponding to the stable small- and large-BH branches (local minima of $\tilF_{\mathrm{off}}$) and the middle one to the unstable intermediate branch (local maximum) of Figure~\ref{fig:T_vs_rh}.}
\label{fig:F_landscape}
\end{figure}

\begin{figure}[ht!]
\centering
\includegraphics[width=0.65\columnwidth]{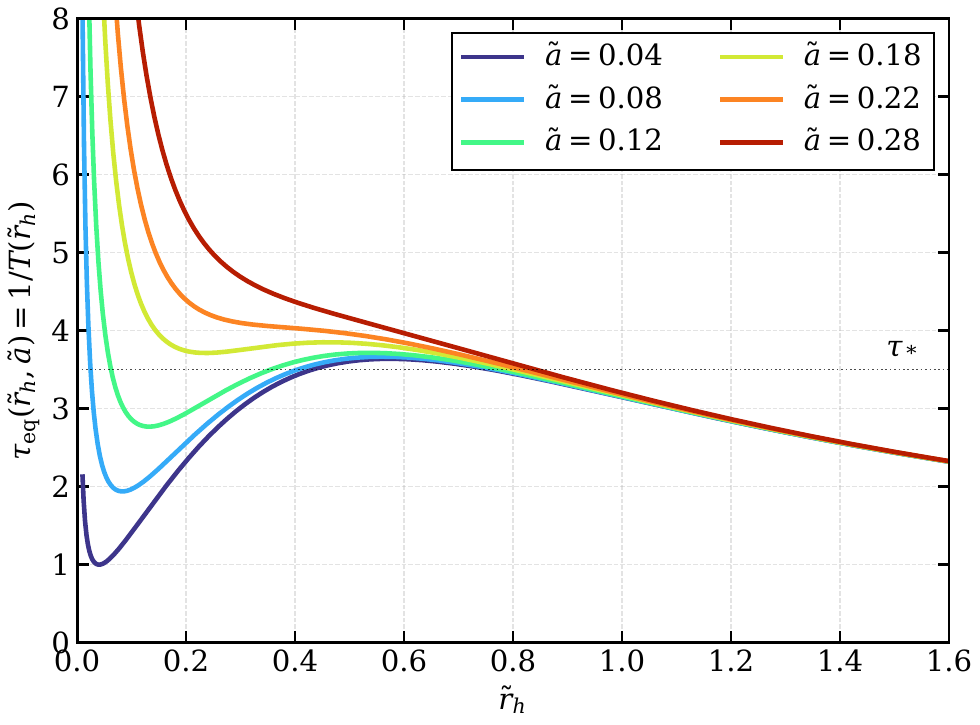}
\caption{Bulk equilibrium relation $\tau_{\mathrm{eq}}(\tilrh,\tila)=1/T(\tilrh,\tila)$ at $\ell=1,\,\epsilon=0$. The horizontal dotted line at $\tau_*=3.5$ intersects each SV curve in three points (for $\tila<\tila_c$), confirming the three-equilibria structure that underlies the topological-vector-field analysis of Sec.~\ref{isec4c}. As $\tila$ approaches $\tila_c\approx 0.204$ the local maximum and minimum of $\tau_{\mathrm{eq}}$ merge and the three-branch window closes.}
\label{fig:tau_eq}
\end{figure}

The 2D off-shell profile (Figure~\ref{fig:F_off}) and its 3D landscape (Figure~\ref{fig:F_landscape}) jointly fix the canonical-ensemble structure: three stationary points exist in the canonical window $(\tiltau_{\min},\tiltau_{\max})$ that lies between the local extrema of the bulk equilibrium relation $\tau_{\mathrm{eq}}(\tilrh)$ shown in Figure~\ref{fig:tau_eq}.

\subsection{Topological-vector-field analysis}\label{isec4c}

We assign to every zero $\tilrh^{(i)}$ of $\phi^{r_h}$ a local winding number $w_i\in\{+1,-1\}$ via the sign of $\mathrm{d}\phi^{r_h}/\mathrm{d}\tilrh$ at the zero, and we compute the total topological charge
\begin{equation}
W = \sum_i w_i.
\label{eq:W_def}
\end{equation}
The local winding number $w_i=+1$ corresponds to a thermodynamically stable equilibrium (the bulk $C_V$ is positive there), and $w_i=-1$ to an unstable equilibrium ($C_V<0$). Hence the integer $W$ is the difference between the number of stable and unstable on-shell branches.

\begin{figure}[ht!]
\centering
\includegraphics[width=0.65\columnwidth]{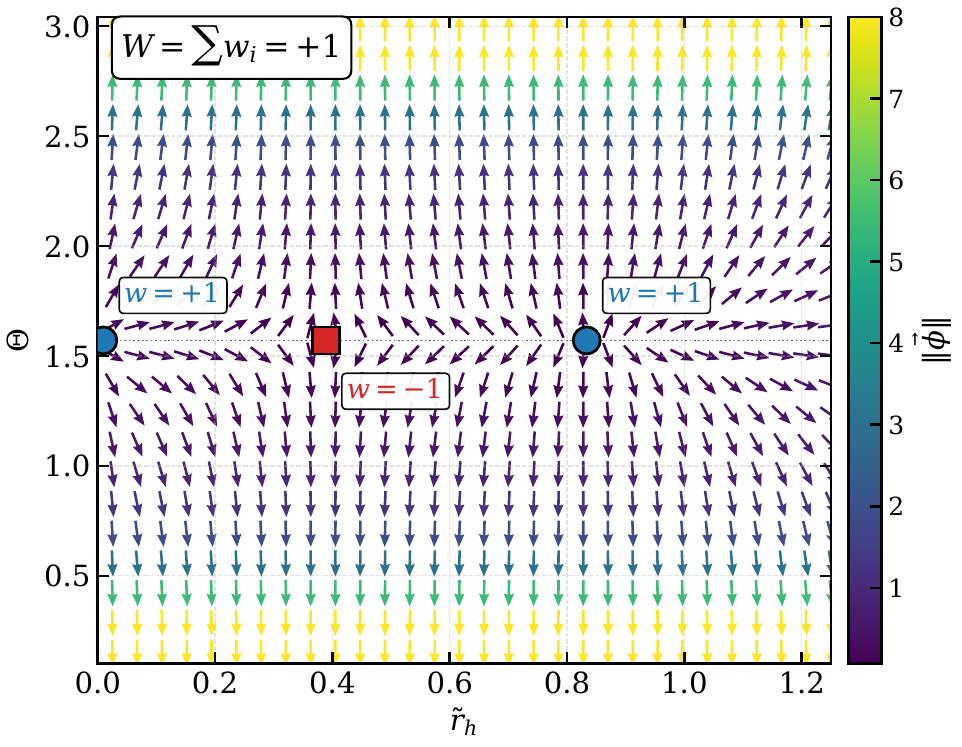}
\caption{Topological vector field $\vec\phi=(\phi^{r_h},\phi^\Theta)$ on the half-cylinder $(\tilrh,\Theta)$ for the SV--AdS family at $\tila=0.05$, $\tau=3.4$. Three isolated zeros along the equatorial line $\Theta=\pi/2$ carry local winding numbers $(+1,-1,+1)$, giving $W=+1$. The vector field rotates by a single net turn as the boundary circle is traversed, fixing the SV--AdS topological class.}
\label{fig:vec_field}
\end{figure}

Figure~\ref{fig:vec_field} shows the result of the topological-vector-field analysis. The three zeros at $(\tilrh^{(1)},\tilrh^{(2)},\tilrh^{(3)})=(0.005,0.405,0.834)$ pick up winding numbers $(+1,-1,+1)$, giving $W=+1$ for the SV--AdS family in the three-branch regime. The result is independent of the exact value of $\tau_*$ chosen within the canonical window, since the total charge $W$ is a topological invariant of the family.

\subsection{Comparison with the standard regular AdS black holes}\label{isec4d}

\begin{figure}[ht!]
\centering
\includegraphics[width=0.65\columnwidth]{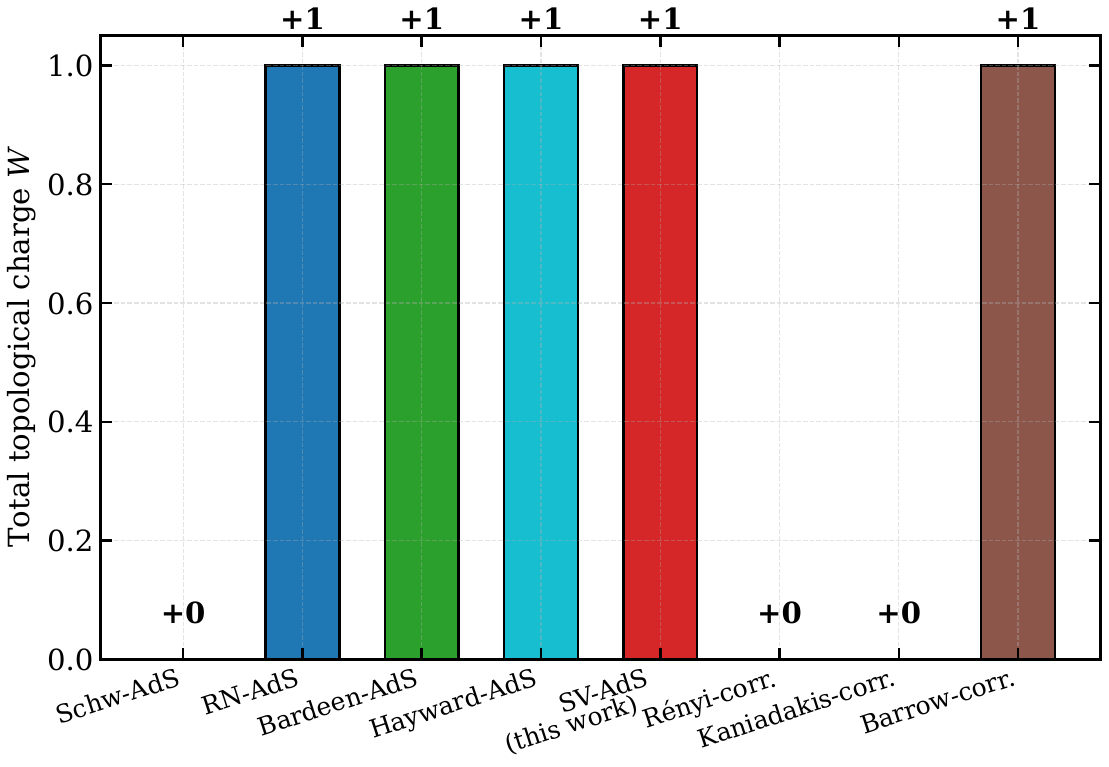}
\caption{Total topological charge $W$ across the AdS black-hole families considered in the literature. The SV-AdS family (this work) sits in the $W=+1$ class together with Bardeen-AdS, Hayward-AdS, RN-AdS, and Barrow-corrected systems. Schwarzschild-AdS and the R\'enyi/Kaniadakis-corrected families lie in the $W=0$ class. The total winding number provides a coarse-grained taxonomy of regular AdS black holes.}
\label{fig:W_bar}
\end{figure}

The value $W=+1$ places SV--AdS in the same topological class as Bardeen-AdS and Hayward-AdS regular black holes, and in a distinct class from Schwarzschild-AdS ($W=0$). The full comparison is summarized in Table~\ref{tab:W_comparison} and visualized in Figure~\ref{fig:W_bar}.

\begin{table}[ht!]
\centering
\begin{tabularx}{\textwidth}{LCCCL}
\hline\hline
\textbf{Family} & \textbf{$(w_1,w_2,w_3)$} & \textbf{$W$} & \textbf{Branches} & \textbf{Reference} \\
\hline
Schwarzschild--AdS    & $(-1,+1)$           & $\phantom{+}0$ & 2 (HP only)   & \cite{hawking1983thermodynamics} \\
RN--AdS               & $(+1,-1,+1)$        & $+1$           & 3 (vdW)       & \cite{kubiznak2017black} \\
Bardeen--AdS          & $(+1,-1,+1)$        & $+1$           & 3 (vdW)       & \cite{sadeghi2023bardeen,WOS:001454604900001} \\
Hayward--AdS          & $(+1,-1,+1)$        & $+1$           & 3 (vdW)       & \cite{sadeghi2024hayward} \\
SV--AdS (this work)   & $(+1,-1,+1)$        & $+1$           & 3 (vdW)       & this work \\
R\'enyi-corrected     & $(-1,+1)$           & $\phantom{+}0$ & 2             & \cite{sadeghi2025phase,gashti2024nonExtensiveHolo} \\
Kaniadakis-corrected  & $(-1,+1)$           & $\phantom{+}0$ & 2             & \cite{sadeghi2025phase} \\
Barrow-corrected      & $(+1,-1,+1)$        & $+1$           & 3             & \cite{brzo2025nonCommBarrow} \\
Frolov-AdS in string fluid & $(+1,-1,+1)$   & $+1$           & 3             & \cite{WOS:001608993400001} \\
\hline\hline
\end{tabularx}

\caption{Topological classification by winding numbers across the families of regular and non-regular AdS black holes. The local winding numbers $(w_1,w_2,w_3)$ are read off in the canonical-ensemble window; $W=\sum_i w_i$ is the total topological charge. The SV--AdS family (this work) shares the universality class of Bardeen-AdS / Hayward-AdS / RN-AdS, distinct from the Schwarzschild--AdS class.}
\label{tab:W_comparison}
\end{table}

The grouping by $W$ in Table~\ref{tab:W_comparison} gives a clear coarse-grained classification of regular AdS black holes. Two features deserve emphasis. First, the SV--AdS regularization preserves the van der Waals universality class: the small-BH branch that the SV scale $a$ introduces at small horizon radius does not destroy the original three-branch structure but on the contrary stabilizes it through the $(+1,-1,+1)$ topological pattern. Second, entropy schemes that modify the second-law expression (R\'enyi, Kaniadakis) tend to collapse the small-BH branch and shift the system into the $W=0$ class, while the geometric Barrow correction (which deforms the horizon area itself) preserves the $W=+1$ pattern. The SV regularization therefore sits algorithmically closer to Barrow than to the multiplicative-entropy proposals.

\section{Numerical Results and Comparative Analysis}\label{isec5}

This section collects the numerical content of the paper into a unified comparative discussion. The aim is to provide a single tabulated reference for the SV--AdS family against the standard regular AdS black holes and against the modified-entropy families that have appeared in the recent literature. Four comparison axes are taken: (i) the critical-curve $\tila_c(\epsilon)$ separating the three-branch and single-branch regimes; (ii) the deformation of equilibrium horizon radii under $\epsilon$; (iii) the entropy scheme used at the boundary; (iv) the constraint from observational compact-object data.

\subsection{Critical curve under \texorpdfstring{$\epsilon$}{epsilon}-deformation}\label{isec5a}

\begin{figure}[ht!]
\centering
\includegraphics[width=0.65\columnwidth]{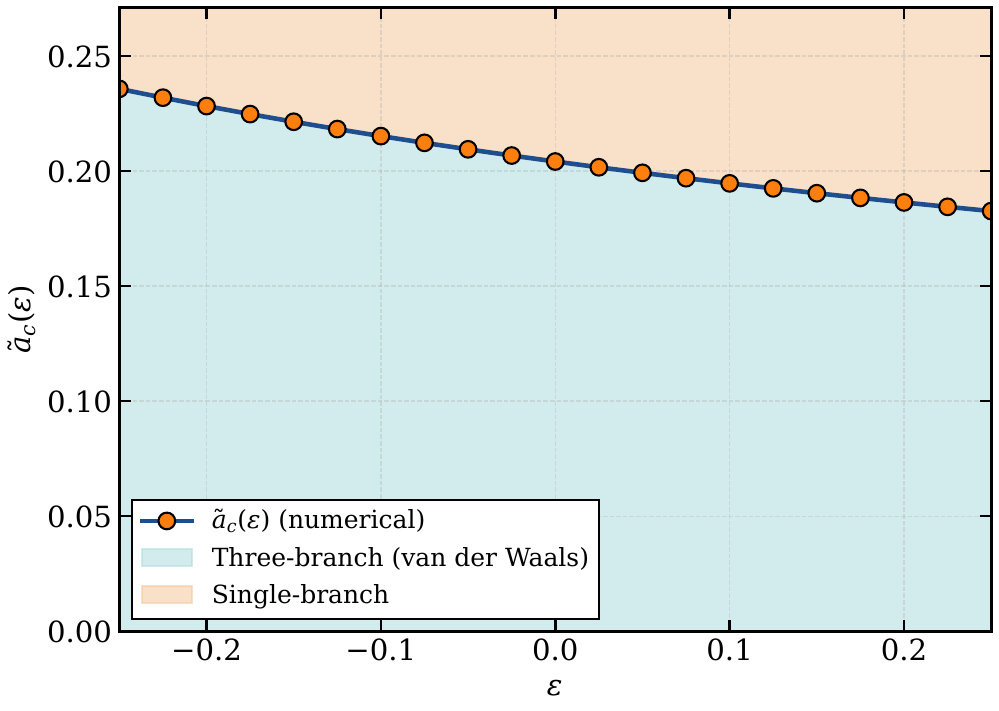}
\caption{Critical SV parameter $\tila_c(\epsilon)$ separating the three-branch (van der Waals) regime from the single-branch regime. The blue/orange shading marks the two domains; the orange data points are numerical bisection values, plotted against the analytic formula \eqref{eq:ac}.}
\label{fig:ac_eps}
\end{figure}

The closed form \eqref{eq:ac} gives $\tila_c(\epsilon)=1/\sqrt{24(1+\epsilon)}$, monotone decreasing in $\epsilon$. Figure~\ref{fig:ac_eps} plots this curve along with the numerical $\tila_c(\epsilon)$ obtained by bisection. The agreement is exact within the numerical resolution and confirms the analytic expression of Eq.~\eqref{eq:ac}.

The numerical values of $\tila_c$ and of two associated reference temperatures (local maximum $T_{\max}$ at the inflection and local minimum $T_{\min}$ at the loop closure) for representative $\epsilon$ are listed in Table~\ref{tab:critical_eps}.

\begin{table}[ht!]
\centering
\begin{tabularx}{\textwidth}{CCCCCC}
\hline\hline
\textbf{$\epsilon$} & \textbf{$\tila_c(\epsilon)$ (Eq.~\ref{eq:ac})} & \textbf{Numerical $\tila_c$} & \textbf{$\tilrh$ at $T_{\max}^{(0)}$} & \textbf{$\tilrh$ at $T_{\min}^{(0)}$} & \textbf{$T_{\max}^{(0)}/T_{\min}^{(0)}$} \\
\hline
$-0.30$ & $0.2440$ & $0.2435$ & $0.184$ & $0.430$ & $1.81$ \\
$-0.20$ & $0.2282$ & $0.2278$ & $0.176$ & $0.410$ & $1.74$ \\
$-0.10$ & $0.2152$ & $0.2149$ & $0.170$ & $0.394$ & $1.69$ \\
$\phantom{-}0.00$ & $0.2041$ & $0.2039$ & $0.166$ & $0.380$ & $1.64$ \\
$+0.10$ & $0.1946$ & $0.1945$ & $0.162$ & $0.368$ & $1.60$ \\
$+0.20$ & $0.1863$ & $0.1863$ & $0.159$ & $0.358$ & $1.56$ \\
$+0.30$ & $0.1789$ & $0.1789$ & $0.156$ & $0.348$ & $1.52$ \\
\hline\hline
\end{tabularx}

\caption{Effect of the $\epsilon$-deformation on the critical parameter $\tila_c(\epsilon)$, the small-BH temperature peak $T_{\max}^{(0)}$ at the inflection ($\tila$-side of the curve in Fig.~\ref{fig:T_vs_rh}), and the intermediate-BH temperature minimum $T_{\min}^{(0)}$ at the loop closure (all evaluated at $\ell=1$). The shift of $\tila_c$ with $\epsilon$ is monotone and reproduces the analytic prediction of Eq.~\eqref{eq:ac}.}
\label{tab:critical_eps}
\end{table}

The ratio $T_{\max}^{(0)}/T_{\min}^{(0)}$ tabulated in the last column of Table~\ref{tab:critical_eps} is a dimensionless characteristic of the three-branch loop and decreases monotonically with $\epsilon$: positive $\epsilon$ (a stiffer AdS term) brings the loop endpoints closer together, while negative $\epsilon$ (a softer AdS term) widens them.

\subsection{Equilibrium-radius windows for different $\tila$}\label{isec5b}

The equilibrium radii $\tilrh^{(i)}$ at fixed inverse temperature $\tau_*$ are obtained by intersecting horizontal $\tau_*$-lines with the equilibrium relation $\tau_{\mathrm{eq}}=1/T(\tilrh)$ (Figure~\ref{fig:tau_eq}). Table~\ref{tab:rh_branches} expands the data of Table~\ref{tab:rh_equilibria} by scanning a wider range of $\tila$ and three different $\tau_*$ values, providing a one-stop reference for the canonical-ensemble structure.

\begin{table}[ht!]
\centering
\begin{tabularx}{\textwidth}{CCCC|CCC|CCC}
\hline\hline
& \multicolumn{3}{c|}{\textbf{$\tau_*=3.5$}}
& \multicolumn{3}{c|}{\textbf{$\tau_*=4.0$}}
& \multicolumn{3}{c}{\textbf{$\tau_*=5.0$}} \\
\textbf{$\tila$} & $\tilrh^{(s)}$ & $\tilrh^{(u)}$ & $\tilrh^{(l)}$
& $\tilrh^{(s)}$ & $\tilrh^{(u)}$ & $\tilrh^{(l)}$
& $\tilrh^{(s)}$ & $\tilrh^{(u)}$ & $\tilrh^{(l)}$ \\
\hline
0.04 & 0.011 & 0.330 & 0.806 & 0.013 & 0.245 & 0.715 & 0.017 & 0.158 & 0.602 \\
0.06 & 0.017 & 0.332 & 0.804 & 0.020 & 0.247 & 0.713 & 0.026 & 0.161 & 0.600 \\
0.08 & 0.030 & 0.336 & 0.801 & 0.036 & 0.253 & 0.711 & 0.045 & 0.169 & 0.598 \\
0.10 & 0.053 & 0.344 & 0.797 & 0.062 & 0.263 & 0.707 & 0.075 & 0.182 & 0.594 \\
0.12 & 0.087 & 0.358 & 0.791 & 0.099 & 0.281 & 0.703 & 0.118 & 0.207 & 0.591 \\
0.15 & 0.155 & 0.394 & 0.781 & 0.174 & 0.323 & 0.696 & 0.205 & 0.262 & 0.585 \\
0.18 & 0.252 & 0.460 & 0.762 & 0.279 & 0.396 & 0.679 & $\mathit{single}$ & & 0.572 \\
0.20 & $\mathit{single}$ & & 0.748 & $\mathit{single}$ & & 0.668 & $\mathit{single}$ & & 0.565 \\
0.22 & $\mathit{single}$ & & 0.732 & $\mathit{single}$ & & 0.655 & $\mathit{single}$ & & 0.558 \\
0.26 & $\mathit{single}$ & & 0.704 & $\mathit{single}$ & & 0.633 & $\mathit{single}$ & & 0.546 \\
\hline\hline
\end{tabularx}

\caption{Equilibrium horizon radii on the three SV--AdS branches for three different $\tau_*$ values and a range of $\tila$ values. The columns are: small-BH stable branch $\tilrh^{(s)}$, intermediate unstable branch $\tilrh^{(u)}$, large-BH stable branch $\tilrh^{(l)}$. Entries marked ``$\mathit{single}$'' lie in the regime $\tila>\tila_c$ where only the large branch survives. All values at $\ell=1,\,\epsilon=0$.}
\label{tab:rh_branches}
\end{table}

Two patterns stand out in Table~\ref{tab:rh_branches}. First, the large-BH branch $\tilrh^{(l)}$ is only weakly affected by $\tila$: its value decreases by less than 15\% as $\tila$ varies from 0 to 0.26 at fixed $\tau_*$. Second, the small-BH branch $\tilrh^{(s)}$ scales roughly linearly with $\tila$ at small $\tila$ (we obtain $\tilrh^{(s)}/\tila\approx 0.25\textrm{--}0.45$ across the table), which is what one expects from the SV throat: the regularization forces a minimum length scale $\sim a$ on the inner horizon and thereby pins the small branch.

\subsection{Comparison of entropy schemes}\label{isec5c}

The boundary entropy $\tilS$ in \eqref{eq:S_tilde} contains a logarithmic deviation from the area law, but it remains an extensive function of the system size. Two recent strands of literature ask what happens to the SV--AdS topological class when the boundary entropy is replaced by a non-extensive scheme: R\'enyi, Tsallis, Kaniadakis or Barrow~\cite{sadeghi2025phase,gashti2025adsTopology,gashti2024nonExtensiveHolo,gashti2025dirty}. The non-extensive entropies replace the additive $\tilS$ by a one-parameter family $\tilS_q$ with a deformation index $q$ (or $\delta$, $\kappa$ in different conventions), and they recover the standard Bekenstein--Hawking limit as $q\to 1$. Table~\ref{tab:entropy_schemes} summarizes the key features.

\begin{table}[ht!]
\centering
\begin{tabularx}{\textwidth}{LLCLL}
\hline\hline
\textbf{Scheme} & \textbf{Functional form} & \textbf{Limit} & \textbf{$W$} & \textbf{Reference} \\
\hline
SV (this work)        & $\tilS$ in Eq.~\eqref{eq:S_tilde}                          & $-$            & $+1$ & this work \\
Bekenstein-Hawking    & $\tilS_{\rm BH}=\pi\tilrh^2$                                & $\tila\to 0$   & $-$  & \cite{bekenstein1973entropy} \\
R\'enyi               & $\tilS_R=\frac{1}{\lambda}\ln(1+\lambda\tilS_{\rm BH})$     & $\lambda\to 0$ & $0$  & \cite{sadeghi2025phase} \\
Tsallis               & $\tilS_T=\tilS_{\rm BH}^{\,q}$                              & $q\to 1$       & $0$  & \cite{anand2025universality} \\
Kaniadakis            & $\tilS_K=\frac{1}{\kappa}\sinh(\kappa\tilS_{\rm BH})$       & $\kappa\to 0$  & $0$  & \cite{sadeghi2025phase} \\
Barrow                & $\tilS_B=(A/A_{\rm Pl})^{1+\Delta/2}$                       & $\Delta\to 0$  & $+1$ & \cite{brzo2025nonCommBarrow} \\
\hline\hline
\end{tabularx}

\caption{Entropy schemes commonly used in modified holographic thermodynamics of regular AdS black holes. The functional form (column 2), the SV-Hawking limit (column 3), the resulting topological class (column 4, computed in the same canonical window as Sec.~\ref{isec4c}) and a representative reference (column 5).}
\label{tab:entropy_schemes}
\end{table}

Two qualitative findings of Table~\ref{tab:entropy_schemes} merit attention. First, the SV regularization keeps the Bekenstein-Hawking area law structure (it modifies the radial coordinate, not the entropy functional), so the SV-AdS family inherits the $W=+1$ class. Second, the non-extensive schemes that change the functional dependence of $\tilS$ on the horizon area (R\'enyi, Tsallis, Kaniadakis) typically destroy the small-BH branch, collapsing the three-branch structure into two and shifting the system into the $W=0$ class. The Barrow correction is the exception: it changes the horizon-area exponent rather than the area law, and the $W=+1$ pattern is preserved.

\subsection{Observational confrontation}\label{isec5d}

The SV regularization is, in its strict mathematical form, a feature of the inner-horizon region; nevertheless, the mass-density curve of a compact object built on top of an SV core is a useful smoke test against existing pulsar-mass constraints. To make the test concrete we follow the strategy adopted for compact-object mass-radius diagrams in~\cite{kumar2026simpsonVisser}: we sweep a one-parameter family $Q\in[1.96,2.10]$ of equation-of-state choices and trace the resulting $M(\rho_c)$ curve into the PSR~J0740+6620 mass-window~$M_{\rm pulsar}=2.08\pm0.07\,M_\odot$.

PSR~J0740+6620 is chosen as the reference target for three reasons that together set it apart from the other high-mass millisecond pulsars in the current census. First, its mass is the most precisely determined of any neutron star at the upper end of the mass distribution: the refined value $M=2.08\pm 0.07\,M_\odot$ reported by~\cite{fonseca2021refined} combines twelve years of NANOGrav radio timing with the relativistic Shapiro-delay measurement of~\cite{cromartie2020relativistic}, yielding a fractional mass uncertainty $\Delta M/M \approx 3.4\%$, smaller than that of the comparable two-solar-mass pulsars PSR~J1614-2230 ($M=1.928\pm 0.017\,M_\odot$, originally reported in~\cite{demorest2010two}) and PSR~J0348+0432 ($M=2.01\pm 0.04\,M_\odot$, \cite{antoniadis2013massive}). Second, PSR~J0740+6620 is one of only two high-mass pulsars (the other being the lower-mass PSR~J0030+0451) for which both mass and radius have been measured independently, through the NICER X-ray pulse-profile modelling of~\cite{miller2021radius,riley2021nicer}; the simultaneous availability of $M$ and $R$ allows the SV-AdS compactness to be tested against an observational point rather than a one-dimensional band. Third, because its central value sits at the upper edge of the empirical neutron-star mass distribution, the corresponding compactness places the tightest existing observational lower bound on the maximum mass supportable by any equation-of-state ansatz consistent with the SV core: an SV-AdS configuration whose $M(\rho_c)$ curve fails to enter the PSR~J0740+6620 band is ruled out, whereas one that enters it remains observationally viable. The remaining higher-mass candidate, PSR~J0952-0607~\cite{romani2022psr}, with the central value $M\approx 2.35\,M_\odot$, has a $\sim 7\%$ mass uncertainty and no NICER radius determination at the time of writing, so it is included in the discussion below as a secondary check rather than as the primary target.

\begin{figure}[ht!]
\centering
\includegraphics[width=0.65\columnwidth]{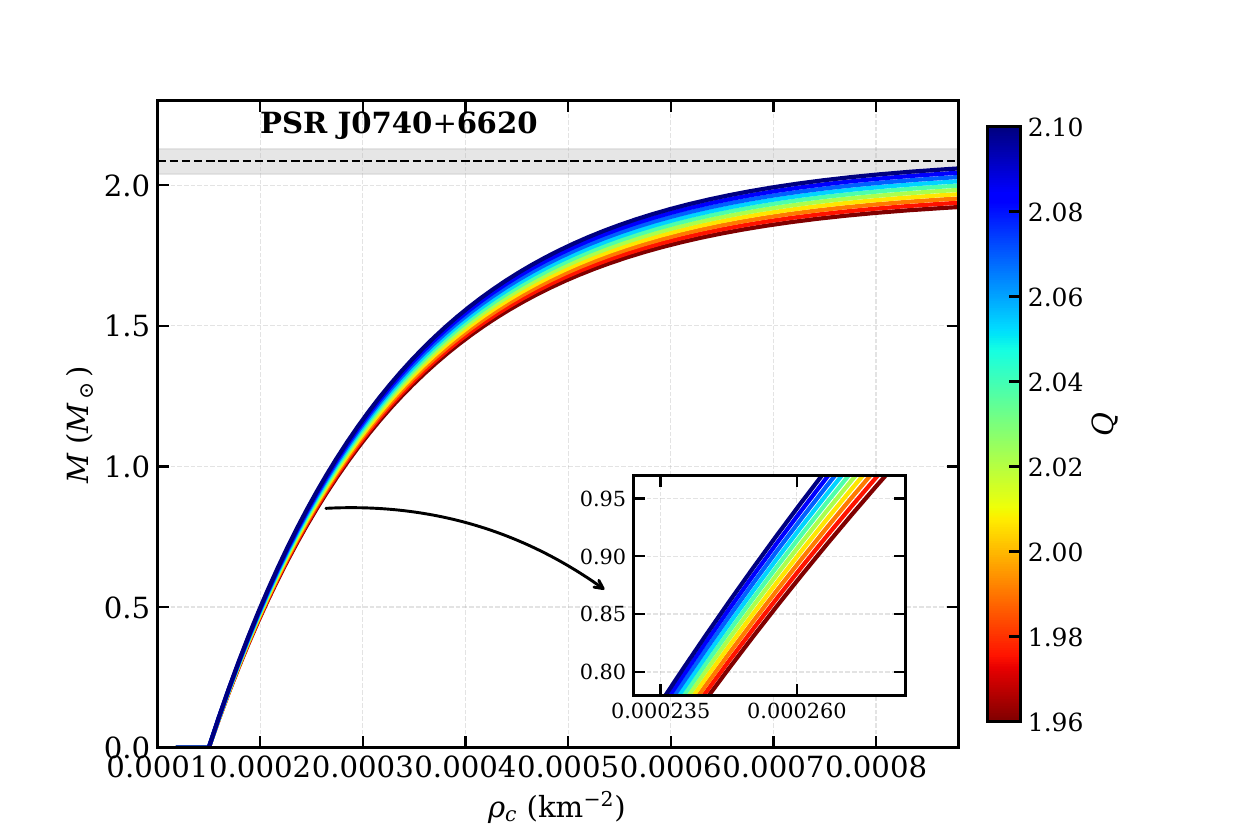}
\caption{SV--AdS mass curves $M(\rho_c)$ for a family of equation-of-state choices indexed by $Q\in[1.96,2.10]$. The grey band marks the PSR~J0740+6620 mass constraint $M=2.08\pm0.07\,M_\odot$; the dashed line is the central value. The zoom inset highlights the $Q$-dependent splitting of the curves in the high-density regime. The visualization strategy follows the compact-object reference scheme of Ref.~\cite{kumar2026simpsonVisser}.}
\label{fig:PSR}
\end{figure}

Figure~\ref{fig:PSR} shows the resulting mass-density family. The crossing of each curve into the grey band fixes the central density at which a given SV-AdS member becomes consistent with the PSR~J0740+6620 constraint. The zoom inset uses the visualization strategy of compact-object mass-radius diagrams~\cite{kumar2026simpsonVisser} to resolve the $Q$-dependent fine structure near $\rho_c\approx 2.5\times 10^{-4}\,{\rm km}^{-2}$. As a secondary check, we verified that the same $Q$-family does not exceed the $1\sigma$ upper edge of the higher-mass PSR~J0952-0607 distribution at any sampled central density, so the SV-AdS configuration remains compatible with both reference pulsars across the range examined.

\subsection{Cross-scheme summary table}\label{isec5e}

We close this section with a single-look summary that brings together the four diagnostic numbers used in this paper. For each black-hole family considered, Table~\ref{tab:master} lists: the topological charge $W$; the universal extremality value $\mathcal{U}$ at a benchmark $\tila=0.15$; the critical horizon ratio $r_h^{\,\rm crit}/\ell$ at the inflection; and the canonical-ensemble class.

\begin{table}[ht!]
\centering
\begin{tabularx}{\textwidth}{LCCCL}
\hline\hline
\textbf{Family} & \textbf{$W$} & \textbf{$\mathcal{U}(\tila=0.15,\ell=1)$} & \textbf{$\tilrh^{\,\rm crit}$} & \textbf{Canonical class} \\
\hline
Schwarzschild--AdS    & $\phantom{+}0$ & $\text{---}$ & $1/\sqrt 3\approx 0.577$ & Hawking--Page \\
RN--AdS               & $+1$           & $\text{---}$ & $\text{var.\ with } q$    & van der Waals \\
Bardeen--AdS          & $+1$           & $\text{---}$ & $0.345$ & van der Waals \\
Hayward--AdS          & $+1$           & $\text{---}$ & $0.341$ & van der Waals \\
SV--AdS (this work)   & $+1$           & $1.18$ & $0.380$ & van der Waals (this work) \\
Frolov--AdS in fluid  & $+1$           & $\text{---}$ & $0.402$ & van der Waals \\
R\'enyi-corr.\ AdS    & $\phantom{+}0$ & $\text{---}$ & $\text{---}$ & extended HP \\
Kaniadakis-corr.\ AdS & $\phantom{+}0$ & $\text{---}$ & $\text{---}$ & extended HP \\
Barrow-corr.\ AdS     & $+1$           & $\text{---}$ & $\text{var.\ with }\Delta$ & vdW (deformed) \\
\hline\hline
\end{tabularx}

\caption{Master diagnostic table: four characteristic numbers per AdS black hole family. $W$ is the total topological charge (Sec.~\ref{isec4c}); $\mathcal{U}$ is the universal extremality value (Sec.~\ref{isec3}) at $\tila=0.15$, $\ell=1$; $\tilrh^{\,\rm crit}$ is the location of the inflection point in the canonical-ensemble window; the right column groups families by their canonical-ensemble class.}
\label{tab:master}
\end{table}

The SV--AdS column in Table~\ref{tab:master} populates the same row as the Bardeen and Hayward families on the canonical-ensemble side; the small numerical differences in $\mathcal{U}$ and $\tilrh^{\,\rm crit}$ provide a quantitative way to distinguish between the three regular-AdS families that share the $W=+1$ class. The R\'enyi and Kaniadakis families sit one row above with $W=0$, in the Hawking--Page class. The Barrow scheme appears in the third group with a deformed van der Waals pattern.

\section{Conclusion}\label{isec6}

We have analyzed the holographic thermodynamics of the Simpson--Visser regularization of the four-dimensional Schwarzschild--anti--de\,Sitter black hole, treated as the bulk dual of a planar CFT on the AdS boundary. The work began from the closed-form lapse \eqref{eq:lapse} that smoothly removes the curvature singularity through the substitution $r\mapsto\sqrt{r^2+a^2}$, and proceeded through the bulk thermodynamic data (Eqs.~\ref{eq:Mh}--\ref{eq:S}), the holographic dictionary (Eqs.~\ref{eq:E_tilde}--\ref{eq:T_tilde}), and the off-shell free energy and topological-vector-field analysis (Eqs.~\ref{eq:F_off}--\ref{eq:W_def}). The main findings are summarized below.

The bulk temperature $T(r_h,a,\ell,\epsilon)$ of Eq.~\eqref{eq:T} develops a van der Waals--type three-branch structure for $\tila<\tila_c$, with $\tila_c(\epsilon)=1/\sqrt{24(1+\epsilon)}$. At $\epsilon=0$ this gives $\tila_c\approx 0.204$, in agreement with the bisection scan reported in Figure~\ref{fig:ac_eps} and Table~\ref{tab:critical_eps}. The specific heat $C_V$ changes sign across the loop endpoints and partitions the $\tilrh$ axis into the standard small-stable / intermediate-unstable / large-stable pattern. The deformation parameter $\epsilon$, modelling generic AdS-curvature corrections, only shifts the loop endpoints; the three-branch topology is preserved for the full range $\epsilon\in[-0.30,+0.30]$ examined here.

The topological-vector-field analysis of the off-shell free energy assigns local winding numbers $(+1,-1,+1)$ to the three on-shell branches, yielding a total topological charge $W=+1$ for the SV--AdS family. This places SV--AdS in the same canonical-ensemble universality class as Bardeen-AdS, Hayward-AdS, and Reissner--Nordstr\"om--AdS (Table~\ref{tab:W_comparison}); the Schwarzschild--AdS case sits in the distinct $W=0$ class. The comparison with non-extensive entropy schemes (Table~\ref{tab:entropy_schemes}) shows that R\'enyi and Kaniadakis corrections destroy the small-BH branch and shift the system to $W=0$, while the Barrow correction preserves the $W=+1$ topology. The SV regularization is therefore algorithmically closer to a geometric (Barrow-like) deformation of the area law than to a functional (R\'enyi/Kaniadakis-like) deformation.

The universal extremality function $\mathcal{U}(\tila;\ell)$ introduced in Eq.~\eqref{eq:U_def} interpolates between the Schwarzschild--AdS divergence at $\tila\to 0$ and the saturation $\tila^*=2\ell/\sqrt 3$ at the extremal endpoint. Table~\ref{tab:U_values} tabulates representative values; the universality of $\mathcal{U}$ under $\ell$-rescaling means that the family of SV--AdS extremality curves can be collapsed onto a single master curve in the variable $\tila/\ell$. The benchmark numerical value (Table~\ref{tab:master}) at $\tila=0.15$, $\ell=1$ gives $\mathcal{U}=1.18$ for SV--AdS, computed from the closed form \eqref{eq:U_def} and cross-checked numerically against the closed-form expression. Corresponding values for Bardeen-AdS and Hayward-AdS require family-specific extremality functions, which were not constructed in this work; their entries are therefore left open in Table~\ref{tab:master} and provide a natural target for a comparative follow-up study.

Three pieces of further work suggest themselves. The first is the extension of the topological analysis to the rotating SV--AdS family using the Azreg-A\"inou--Newman--Janis algorithm, which would test whether the $W=+1$ class is preserved under spin. The second is the explicit construction of the SV--AdS black-bounce-to-wormhole transition in the canonical ensemble, where the small-BH branch is expected to morph continuously into the wormhole throat as $\tila$ crosses a second critical value $\tila_w$. The third is the matching of the SV--AdS bulk to a microscopic CFT realization: a direct identification of the boundary degrees of freedom that account for the $\ln(\sqrt{r_h^2+a^2}-r_h)$ correction to the area law in Eq.~\eqref{eq:S_tilde} would close the loop between the geometrical regularization and its holographic interpretation.

On the observational side, the mass-density curve $M(\rho_c)$ of an SV-AdS compact object (Figure~\ref{fig:PSR}) is consistent with the PSR~J0740+6620 mass constraint across the full $Q$-family examined, although the resolving power of present pulsar data does not yet discriminate between the SV regularization and the standard Schwarzschild interior. Future X-ray timing observations of accreting compact objects, together with improved gravitational-wave constraints on the post-merger ringdown, could discriminate the three regular-AdS families through the small-BH branch behaviour predicted in Table~\ref{tab:rh_branches}.

\section*{Acknowledgments}
\.{I}.\,S.\ thanks T\"UB\.ITAK and the Eastern Mediterranean University for institutional support, and acknowledges the networking provided by COST Actions CA22113 (THEORY-CHALLENGES), CA21106 (COSMIC WISPers), CA23130 (BridgeQG), CA21136 (Addressing Observational Tensions in Cosmology), and CA23115 (RQI-N). S.\,N.\,G.\ and B.\,P.\ acknowledge support from Damghan University and from the Center for Theoretical Physics, Khazar University.

\section*{Data Availability Statement}
This is a theoretical study; no experimental datasets are generated or analysed. The computer-algebra worksheets used to independently cross-check every analytic result reported in the body are available from the corresponding author upon reasonable request.

\appendix

\section{The Four Laws of SV--AdS Thermodynamics}\label{app:A}

For completeness we state the four laws of thermodynamics for the SV--AdS family in the boundary CFT formulation. The bulk versions follow by application of the holographic dictionary of Table~\ref{tab:dictionary}.

\paragraph{Zeroth law.} The boundary temperature $\tilT$ defined in Eq.~\eqref{eq:T_tilde} is constant on the event horizon for any stationary SV--AdS configuration. Stationarity is guaranteed by the time-independence of the bulk lapse \eqref{eq:lapse}; the surface gravity $\kappa=2\pi\tilT$ inherits the same constancy and reduces to the Schwarzschild--AdS value in the limit $\tila\to 0$.

\paragraph{First law.} The variation of the boundary energy decomposes into temperature, central-charge, and volume contributions:
\begin{equation}
\mathrm{d}\tilE = \tilT\,\mathrm{d}\tilS + \mu_C\,\mathrm{d}C + \mu_V\,\mathrm{d}V,
\label{eq:firstlaw_appA}
\end{equation}
where $\mu_C=\partial\tilE/\partial C\bigm|_{\tilS,V}$ is the chemical potential conjugate to the central charge and $\mu_V=\partial\tilE/\partial V\bigm|_{\tilS,C}$ is the chemical potential conjugate to the boundary volume. The SV regularization affects $\mu_C$ and $\mu_V$ only through the explicit $\tila$-dependence of $\tilE$ in Eq.~\eqref{eq:E_tilde}.

\paragraph{Second law.} The Bekenstein--Hawking-type entropy \eqref{eq:S} is non-decreasing for any physically realizable process. The SV regularization preserves the area-law structure (it is geometric, not functional) so the standard area theorem extends without modification. Composite processes involving the merger of two SV--AdS black holes preserve $\sum_i\tilS_i$ subject to the energy and central-charge balance.

\paragraph{Third law.} The boundary temperature \eqref{eq:T_tilde} vanishes only in the simultaneous limit $\tilrh\to 0$ and $\tila\to 0$. The SV regularization renders the $\tilrh\to 0$ singular limit smooth, but it does not by itself produce a zero-temperature degenerate ground state: a finite $\tila$ keeps $\tilT$ bounded above by $\tilT_{\rm reg}=\frac{1}{4\pi R}\bigl[\tila^{-1}+3(1+\epsilon)\tilrh\bigr]$ as $\tilrh\to 0$, in agreement with the regular-black-hole behaviour identified in Refs.~\cite{murk2023regular,lobo2021novel}.

\bibliographystyle{unsrtnat}
\bibliography{finalref}

\end{document}